\documentclass{ws-mplb}

\begin{document}

%\markboth{Authors' Names}{Instructions for  
%Typing Manuscripts (Paper's Title)}

\catchline{}{}{}{}{}
%
%%%%%%%%%%%%%%%%%%%%%%%%%%%%%%%%%%%%%%%%%%%%%%%%%%%%%%%%%%%%%%%%%%%%

\title{FREE EXPANSION OF IMPENETRABLE BOSONS ON 
\\ ONE-DIMENSIONAL OPTICAL LATTICES}

\author{\footnotesize MARCOS RIGOL}
\address{Physics Department, University of California, Davis,
CA 95616, USA}
\author{ALEJANDRO MURAMATSU}
\address{Institut f\"ur Theoretische Physik III, Universit\"at 
Stuttgart, 70550 Stuttgart, Germany}

\maketitle

\begin{history}
\received{(Day Month Year)}
\revised{(Day Month Year)}
\end{history}

\begin{abstract}
We review recent exact results for the free expansion 
of impenetrable bosons on one-dimensional lattices, after 
switching off a confining potential. When the system is 
initially in a superfluid state, far from the regime in 
which the Mott-insulator appears in the middle of the trap, 
the momentum distribution of the expanding bosons rapidly 
approaches the momentum distribution of noninteracting fermions. 
Remarkably, 
no loss in coherence is observed in the system as reflected 
by a large occupation of the lowest eigenstate of the one-particle 
density matrix. In the opposite limit, when the initial system 
is a pure Mott insulator with one particle per lattice site, 
the expansion leads to the emergence of quasicondensates at 
finite momentum. In this case, one-particle correlations 
like the ones shown to be universal in the equilibrium case
develop in the system. We show that the out-of-equilibrium 
behavior of the Shannon information entropy in momentum 
space, and its contrast with the one of noninteracting fermions, 
allows to differentiate the two different regimes of interest. 
It also helps in understanding the crossover between them.
\end{abstract}


\section{Introduction}

Physical systems displaying the effects of a reduced dimensionality 
have been increasingly attracting the interest of experimental and 
theoretical physicists in the past decades. Traditionally a 
condensed-matter subject, it has received a lot of attention 
within the last years in the framework of ultracold quantum gases.
The realization of Bose-Einstein condensates (BEC) in very anisotropic 
traps,\cite{schreck01,gorlitz01,meyrath05} or their loading on optical
lattices,\cite{greiner01,moritz03,stoferle04,tolra04,paredes04,kinoshita04,fertig05}
allow experimentalists to obtain a rich variety of (quasi-)one-dimensional 
(1D) systems where the reduced dimensionality rules the physics. 

In particular, the 1D regime where bosons behave as 
impenetrable particles,\cite{paredes04,kinoshita04} also known as hardcore 
bosons (HCB's),  has generated much interest recently. 
This regime, which has been called the Tonks-Girardeau (TG) 
regime,\cite{girardeau60} can be obtained for large positive 
three-dimensional scattering lengths, low densities, and low 
temperatures.\cite{olshanii98,petrov00,dunjko01} Contrary to weakly
interacting bosons, HCB's in 1D share many properties with noninteracting 
spinless fermions to which they can be mapped.\cite{girardeau60}
(The TG regime is sometimes referred as the fermionized regime.) 
Properties like the energy, density profiles, and density-density correlations, 
are identical in both systems. On the contrary, one-particle correlations, 
and related quantities like the momentum distribution function ($n_k$), 
and the so-called natural orbitals, are very different.

The 1D TG gas, introduced by Girardeau in the
60's,\cite{girardeau60} has been extensively studied in the literature 
by means of different techniques, both in 
homogeneous,\cite{lieb63,lenard64,vaidya79,haldane81,korepin93,cazalilla04_1} 
and harmonically 
trapped\cite{girardeau01,papenbrock03,forrester03,gangardt04} systems. 
In 1D, even at zero temperature, HCB's do not 
exhibit BEC. As shown by Lenard,\cite{lenard64} and Vaidya 
and Tracy,\cite{vaidya79} one-particle correlations decay with a 
power law $\rho_x\sim 1/\sqrt{x}$. Hence, the highest 
occupied effective single-particle state (the lowest natural 
orbital) in the system has an occupation 
$\lambda_0\sim \sqrt{N_{b}}$ ($N_b$ being the number 
of bosons), i.e., $\lambda_0\rightarrow\infty$ for 
$N_b\rightarrow\infty$, but $\lambda_0/N_b\rightarrow 0$,
and we call this state a quasicondensate. 
A similar result has been obtained in 
harmonic traps.\cite{forrester03}

Once the 1D regime has been achieved experimentally, for 
example by means of a two-dimensional (2D) optical 
lattice,\cite{greiner01,moritz03,tolra04,kinoshita04} the 
addition of a new lattice along the 1D axes of the 
tubes\cite{paredes04} facilitates the achievement of the TG 
regime with respect to the continuum case. This is because 
it allows experimentalists to change the effective mass of 
the particles, and consequently the ratio between interaction 
and kinetic energies.\cite{paredes04} In 1D lattices the 
HCB Hamiltonian can be mapped onto the 1D XY model of Lieb, 
Schulz, and Mattis.\cite{lieb61} For periodic systems, this 
model has been also studied extensively in the 
literature,\cite{mccoy68,vaidya78,mccoy83} and more 
recently in traps.\cite{rigol04_1} As in the continuum
case, in a periodic lattice the HCB one-particle density 
matrix decays as $\rho_x\sim 1/\sqrt{x}$ for noninteger fillings, 
so that the highest occupied effective single particle state
has an occupation $\sim \sqrt{N_{b}}$. Remarkably, 
the above properties are universal independently of the 
power of the confining potential in the trapped case, 
and of the presence of Mott-insulating domains in the 
system.\cite{rigol04_1}

One fascinating possibility opened by the experimental realization 
of the TG regime is the study of the nonequilibrium 
dynamics of such strongly correlated bosonic systems. As we 
review in this article, two very peculiar features occur during
the expansion of a HCB gas on a 1D lattice. For low initial 
densities, the momentum distribution of expanding HCB's rapidly 
approaches that of noninteracting fermions.\cite{rigol04_3}
This occurs without any evidence of loss of coherence in the 
system, but rather with a slight increment of it, as reflected 
by the occupation of the lowest natural orbital. On the other 
hand, when the system starts its free evolution 
from a pure Mott-insulating (Fock) state, quasicondensates of 
HCB's emerge at finite momentum.\cite{rigol04_2}

We also present in this work a detailed study of the 
out-of-equilibrium behavior of the Shannon information 
entropies\cite{shannon48} in coordinate and momentum space, 
for the two situations mentioned before. As the application 
of information concepts to the study of correlated systems 
is increasing, we find that during the expansion of the TG gas
in a lattice, the Shannon information entropy in momentum space 
displays a qualitative different behavior depending on the 
initial conditions in the system. This behavior allows to 
identify the regime where a fast fermionization of HCB momentum 
distribution occurs, from the one in which traveling 
quasicondensates are dynamically generated in the system,
and study the crossover between them. 

The exposition is organized as follows. In Sec.\ \ref{EA} we 
briefly describe the exact numerical approach we follow to study 
the nonequilibrium dynamics. We also introduce concepts that will 
be used in the rest of the manuscript. In Sec.\ \ref{DF} we 
analyze the dynamical fermionization of the HCB momentum 
distribution function. The emergence of quasicondensates
at finite momentum is reviewed in Sec.\ \ref{EQFM}. Section 
\ref{IE} is devoted to the analysis of the Shannon information 
entropies in coordinate and momentum space for the two cases
discussed in Secs.\ \ref{DF} and \ref{EQFM}, and their contrast
to the ones in the fermionic case. Finally, the conclusions are 
presented in Sec.\ \ref{C}.

\section{Exact Approach \label{EA}}

The HCB Hamiltonian can be written as 
\begin{equation}
\label{HamHCB} H = -t \sum_{i} \left( b^\dagger_{i} b^{}_{i+1}
+ h.c. \right) + V_2 \sum_{i} x_i^2 \ n_{i }, \quad 
b^{\dagger 2}_{i}= b^2_{i}=0, \  
\left\lbrace  b^{}_{i},b^{\dagger}_{i}\right\rbrace =1,
\end{equation}
where the bosonic creation and annihilation operators at site $i$ 
are denoted by $b^{\dagger}_{i}$ and $b^{}_{i}$ respectively, 
and the local density operator by $n_i=b^{\dagger}_{i}b^{}_{i}$.
$t$ is the hopping parameter, and the last term represents a 
harmonic trap with curvature $V_2$ and site positions $x_i=ia$ 
($a$ is the lattice constant). The brackets in Eq.\ (\ref{HamHCB}) 
apply only to on-site anticommutation relations, for $i\neq j$ 
these operators commute  
$[b^{}_{i},b^{\dagger}_{j}]=0$.

To calculate HCB properties we 
use the Jordan-Wigner transformation,\cite{jordan28}
\begin{equation}
\label{JordanWigner} b^{\dag}_i=f^{\dag}_i
\prod^{i-1}_{\beta=1}e^{-i\pi f^{\dag}_{\beta}f^{}_{\beta}},\ \ 
b_i=\prod^{i-1}_{\beta=1} e^{i\pi f^{\dag}_{\beta}f^{}_{\beta}}
f_i \ ,
\end{equation}
which maps the HCB Hamiltonian onto the one of 
noninteracting spinless fermions
\begin{eqnarray}
\label{HamFerm} H_F =-t \sum_{i} \left( f^\dagger_{i}
f^{}_{i+1} + h.c. \right)+ V_2 \sum_{i} x_i^2 \
n^f_{i },
\end{eqnarray}
where $f^\dagger_{i}$ and $f_{i}$ are the creation and
annihilation operators for spinless fermions at site 
$i$, and $n^f_{i}=f^\dagger_{i}f^{}_{i}$ is the local 
particle number operator.

The above mapping allows to express the equal-time Green's function 
for HCB's in a non-equilibrium system as
\begin{equation}
\label{green1} G_{ij}(\tau)=\langle \Psi_{B}(\tau)|
b^{}_{i}b^\dagger_{j}|\Psi_{B}(\tau)\rangle 
=\langle \Psi_{F}(\tau)| \prod^{i-1}_{\beta=1}
e^{i\pi f^{\dag}_{\beta}f^{}_{\beta}} f^{}_i f^{\dag}_j
\prod^{j-1}_{\gamma=1} e^{-i\pi f^{\dag}_{\gamma}f^{}_{\gamma}}
|\Psi_{F}(\tau)\rangle,
\end{equation}
where $\tau$ is the real time variable, $|\Psi^{G}_{B}(\tau)\rangle$ 
is the time evolving wave function for HCB's and 
$|\Psi^{G}_{F}(\tau)\rangle$ is the corresponding one 
for noninteracting fermions.

Since there are no interactions in the equivalent fermionic system, 
the time evolution of an initial wave-function ($|\Psi^I_{F}\rangle$) 
can be calculated numerically as
\begin{equation}
\label{time} 
|\Psi_{F}(\tau)\rangle=e^{-iH_F\tau/\hbar}|\Psi^I_{F}\rangle 
= \prod^{N_f}_{\delta=1}\ \sum^N_{\sigma=1} \ P_{\sigma 
\delta}(\tau)f^{\dag}_{\sigma}\ |0 \rangle,
\end{equation} 
which is a product of single particle states, with $N_f$ being 
the number of fermions ($N_f=N_b$), $N$ the number of lattice sites, 
and ${\bf P(\tau)}$ the matrix of components of $|\Psi_{F}(\tau)\rangle$. 
Once $|\Psi_{F}(\tau)\rangle$ is calculated, the action of 
$\prod^{j-1}_{\gamma=1} e^{-i\pi f^{\dag}_{\gamma}f_{\gamma}}$ 
on the right in Eq.\ (\ref{green1}) generates only a 
change of sign on the elements $P_{\sigma \delta}(\tau)$ 
for $\sigma \leq j-1$, and the further creation of a particle at site 
$j$ implies the addition of one column to ${\bf P(\tau)}$ with the 
element $P_{jN_f+1}(\tau)=1$, and all the others equal to zero. 
The same can be done for the action of $\prod^{i-1}_{\beta=1} e^{i\pi 
f^{\dag}_{\beta}f_{\beta}} f_i$ on the left of Eq.\ (\ref{green1}). 
Hence, the HCB Green's function can be obtained as\cite{rigol04_1}
\begin{eqnarray}
\label{determ}
G_{ij}(\tau)
&=&\langle 0 | \prod^{N_f+1}_{\delta=1}\ \sum^N_{\beta=1} \ 
P'^{A}_{\beta \delta}(\tau)f_{\beta} 
\prod^{N_f+1}_{\sigma=1}\ \sum^N_{\gamma=1} \ P'^{B}_{\gamma 
\sigma}(\tau)f^{\dag}_{\gamma}\ |0 \rangle \nonumber \\
&=&\det\left[ \left( {\bf P}^{'A}(\tau)
\right)^{\dag}{\bf P}^{'B}(\tau)\right],
\end{eqnarray}
where ${\bf P'}^{A}(\tau)$ and ${\bf P'}^{B}(\tau)$ are the new matrices
calculated from ${\bf P}(\tau)$ when the required signs are changed and
the new columns added. The last equality in Eq.\ (\ref{determ}) 
is obtained after considering the following identity
\begin{equation}
\langle 0 |f_{\sigma_1}\cdot \cdot \cdot f_{\sigma_{N_f+1}}
f^{\dag}_{\bar{\sigma}_{N_f+1}} \cdot \cdot \cdot f^{\dag}_{\bar{\sigma}_1}|0 \rangle
=\epsilon^{\lambda_1\cdot \cdot \cdot \lambda_{N_f+1}}
\delta_{\sigma_1\bar{\sigma}_{\lambda_1}}\cdot \cdot \cdot
\delta_{\sigma_{N_f+1}\bar{\sigma}_{\lambda_{N_f+1}}},
\end{equation}
where $\epsilon^{\lambda_1\cdot \cdot \cdot \lambda_{N_f+1}}$ is 
the Levi-Civita symbol in $N_f+1$ dimensions, the indices $\lambda$ 
have values between one and $N_f+1$. 

The evaluation of $G_{ij}(\tau)$ is done numerically, and the equal-time
one-particle density matrix [$\rho_{ij}(\tau)$] is determined by the 
expression 
\begin{equation}
\rho_{ij}(\tau)=\langle \Psi_{B}(\tau)|b^\dagger_{i}b_{j}|\Psi_{B}(\tau)\rangle 
=G_{ji}(\tau)+\delta_{ij}\left[1-2 G_{ii}(\tau)\right].
\end{equation}

There are two quantities related to $\rho_{ij}$ that will be studied 
here because of their relevance to experiments, and to the 
understanding of the coherence of the system. These quantities 
are the momentum distribution function ($n_k$), and the so-called 
natural orbitals. $n_k$ is usually obtained experimentally in 
time-of-flight measurements. We calculate it as
\begin{equation}
n_k(\tau)=(a/\zeta)\sum_{jl} e^{-ik(j-l)}\rho_{jl}(\tau), \quad 
\zeta=\left( V_2/t\right)^{-1/2}.
\end{equation}
Notice that we normalize $n_k$ using $\zeta$, which is a length scale
set by the combination lattice-harmonic confining potential in (\ref{HamHCB}).
Another quantity that is relevant to define by means of $\zeta$ is the 
characteristic density $\tilde{\rho}=N_b/\zeta$.\cite{rigol04_1,rigol03_3}
It plays for trapped systems a role similar to the one of the 
density $\rho=N_b/N$ in the periodic case. As $\tilde{\rho}\to 0$, 
the lattice effects dissappear, and one recovers the continuum limit. 
On the other hand, as $\tilde{\rho}$ increases beyond a critical 
value ($\tilde{\rho}_c\sim 2.6-2.7$) a MI region builds up in the 
middle of the trap.\cite{rigol04_1,rigol03_3}

Also of interest are the natural orbitals ($\phi^\eta$). They can 
be considered as effective single-particle states in these strongly 
interacting systems, and are defined as the eigenfunctions of 
$\rho_{ij}$,\cite{penrose56}
\begin{equation}
\label{NatOrb}
\sum^N_{j=1} \rho_{ij}(\tau)\phi^\eta_j(\tau)=
\lambda_{\eta}(\tau)\phi^\eta_i(\tau),
\end{equation}
having occupations $\lambda_{\eta}$. In dilute higher dimensional 
gases, when only the lowest natural orbital (the highest occupied one) 
scales $\sim N_b$, it can be regarded as the BEC order 
parameter, i.e., the condensate.\cite{leggett01} Although for 
periodic systems in equilibrium the natural orbitals are momentum 
states, this is not in general the case for trapped or 
out-of-equilibrium systems.

\section{Dynamical Fermionization\label{DF}}

\begin{figure}[h]
\begin{center}
\includegraphics[width=0.93\textwidth,height=0.35\textwidth]
{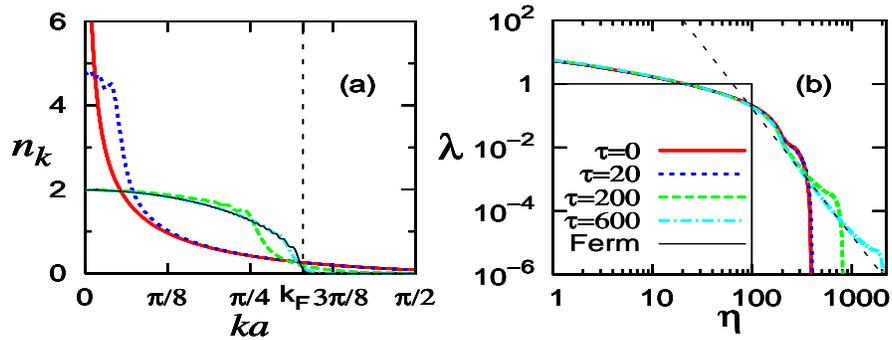}
\end{center} \vspace{-0.44cm}
\caption{$n_k$ (a) and $\lambda$ (b) for 100 HCB's expanding from 
an initial state with $V_2 a^2=2.6\times 10^{-5} t$, and 
compared to the ones for the corresponding fermions. Times ($\tau$) 
are given in units of $\hbar/t$. In (a), $k_F$ denotes the Fermi 
momentum, and $a$ the lattice constant. In (b), the thin dashed 
line corresponds to a power law $\eta^{-4}$, which is known from 
equilibrium systems (see text).}
\label{fermionization}
\end{figure}

In Fig.\ \ref{fermionization}(a) we show the evolution of $n_k$ 
for 100 HCB's once the harmonic trap confining the system 
is turned off. At $\tau=0$, the characteristic density in the 
trap is $\tilde{\rho}=0.51$, i.e., far from the regime 
with a Mott insulator. We compare the HCB $n_k$ with the 
one of the equivalent noninteracting fermions $n^f_k$ 
(which remains unchanged during the expansion). The Fermi
momentum $k_F$ is defined as $\epsilon_F=-2t\cos(k_Fa)$, 
where $\epsilon_F$ is the energy of the last occupied fermionic 
single-particle state in the trap at $\tau=0$. In contrast 
to a periodic system, in a trap $n_k$ is continuous at $k_F$. 
However, it approaches zero even faster than an exponential for $k>k_F$.

Two remarkable features are evident in Fig.\ \ref{fermionization}(a).
(i) Shortly after switching off the trapping potential, the peak at 
$n_{k=0}$ disappears. (ii) The expansion of the system leads to an 
$n_k$ for the HCB's that is equal to the one of the fermions. This 
fermionization of $n_k$ starts from the low momentum states towards 
$k_F$, and produces a Fermi edge.

\begin{figure}[h]
\begin{center}
\includegraphics[width=0.93\textwidth,height=0.35\textwidth]
{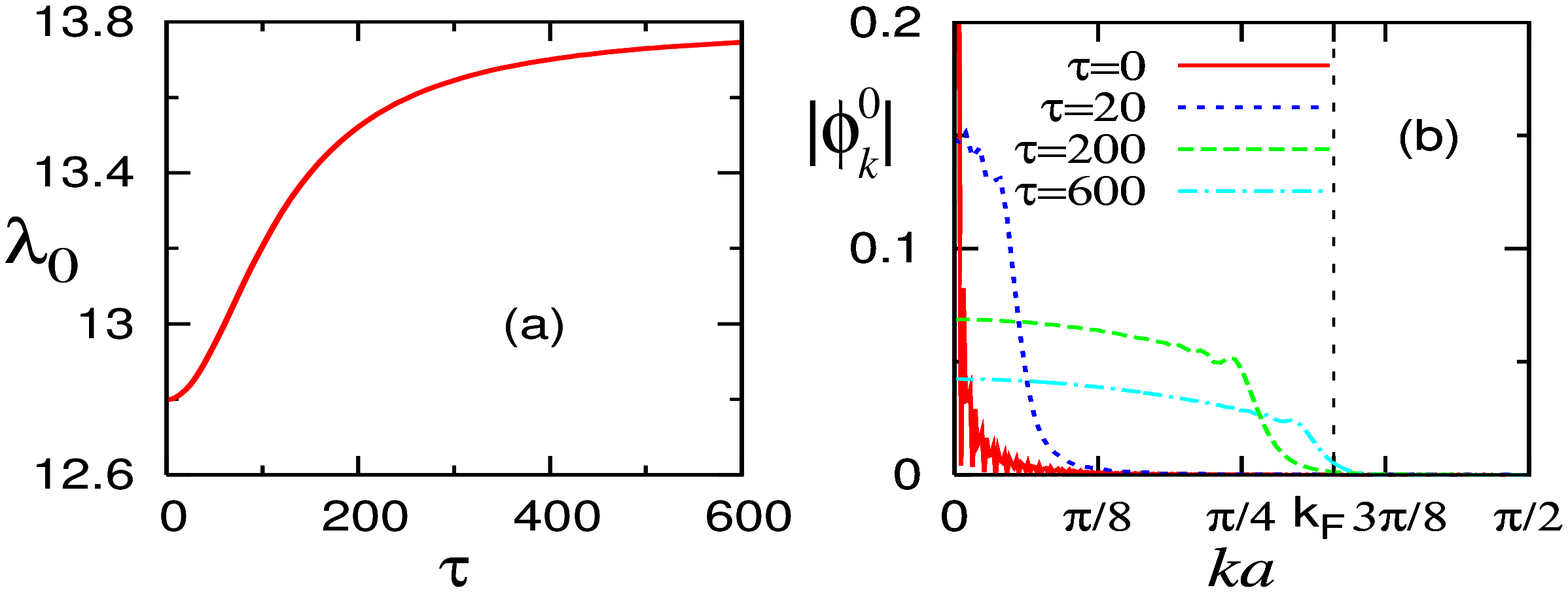}
\end{center} \vspace{-0.44cm}
\caption{Lowest natural orbital occupation (a) and the Fourier transform
of the lowest natural orbital (b) for 100 HCB's expanding from 
an initial state with $V_2 a^2=2.6\times 10^{-5} t$. Times ($\tau$) 
are given in units of $\hbar/t$. In (b), $k_F$ denotes de Fermi momentum.}
\label{NOfermionization}
\end{figure}

After observing the above behavior of the HCB $n_k$, one could 
infer that the system is loosing coherence, i.e.\ that
the effective single-particle states obtained after diagonalizing
$\rho_{ij}$ may reduce their occupations ($\lambda$) with time 
toward $\lambda=1$, the value for noninteracting fermions.
As shown in Fig.\ \ref{fermionization}(b) this is not the case. 
The occupation of the natural orbitals almost does not change. 
Actually, as seen in Fig.\ \ref{NOfermionization}(a) the lowest 
natural orbital occupation $\lambda_0$ slightly increases its 
occupation instead of reducing it.

The out-of-equilibrium increase of $\lambda_0$ is similar to one 
in the ground state when, keeping constant the number of particles, 
the curvature of the trap is decreased.\cite{rigol04_1} In both cases 
the increment of $\lambda_0$ can be intuitively understood as an 
enhancement of the coherence in the system due to an increase of 
its size, which delocalize the HCB's over more lattice sites. However, 
in equilibrium, $n_{k=0}$ also increases along with $\lambda_0$. 
The different behavior of $\lambda_0$ vs $n_{k=0}$ in- and 
out-of-equilibrium can be understood since in the last case the 
lowest natural orbital is composed by HCB's with many different 
momenta. This can be seen in Fig.\ \ref{NOfermionization}(b), 
where we show the Fourier transform of the lowest natural orbital 
($|\phi^0_k|$) at different times. At $\tau=0$ one can see that 
$|\phi^0_k|$ has a peak at $k=0$ showing that quasi-condensation 
occurs around $k=0$, and this is reflected in $n_k$. For $\tau>0$ 
the lowest NO Fourier transform extends in $k$-space so that 
$\phi^0$ starts to be composed by HCB's with low and large momenta, 
i.e., its structure changes in momentum space during the expansion.

In Fig.\ \ref{fermionization}(b) there is a further salient feature, 
which sets in when the density of the expanding HCB's becomes very low. 
A decay $\lambda_\eta\sim \eta^{-4}$ develops for large 
values of $\eta$. In equilibrium we have shown that such a power-law
decay is universal at low densities, independently of the power 
of the confining potential.\cite{rigol04_1} The prefactor of the 
power law $A_{N_b}$ was found to depend only on $N_b$.\cite{rigol04_1}
We find out-of-equilibrium that $A_{N_b}$ has exactly the same value 
than in the ground-state.\cite{rigol04_1} For $N_b=100$ we have 
plotted $\lambda_\eta= A_{N_b}\eta^{-4}$ in 
Fig.\ \ref{fermionization}(b). Presumably, the universal decay 
$\lambda_\eta\sim \eta^{-4}$ is related to the singular character 
of the HCB $\delta$-interaction, as the case for the tail 
$n_k\sim |k|^{-4}$ obtained for continuous 
systems.\cite{minguzzi02,olshanii03} However, analytical proofs of 
this, and the universality of the prefactor $A_{N_b}$, have not been 
given so far. The power law $n_k\sim |k|^{-4}$, that in 
equilibrium\cite{rigol04_1} appears together with 
$\lambda_\eta\sim \eta^{-4}$, disappears during the expansion, 
and the HCB $n_k$ starts to behave like the one of the fermions.

\begin{figure}[h]
\begin{center}
\includegraphics[width=0.94\textwidth,height=0.35\textwidth]
{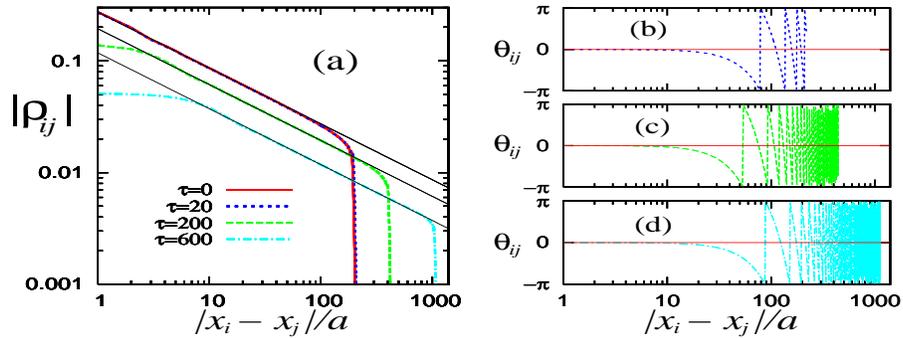}
\end{center} \vspace{-0.44cm}
\caption{Modulus of $\rho_{ij}$ (a) and its phase (b)-(d) 
for the same initial trap parameters and times of Figs.\ 
\ref{fermionization} and \ref{NOfermionization}(b). 
Both quantities have been calculated with respect to the 
center of the system. Thin continuous lines in (a) 
correspond to power laws $|x_i-x_j|^{-1/2}$.}
\label{DensMatrix}
\end{figure}

An overall understanding of the previously discussed nonequilibrium
behavior of $n_k$, and natural orbital occupations, can be gained
directly studying the one-particle density matrix. Out of equilibrium
$\rho_{ij}=|\rho_{ij}|e^{i\theta_{ij}}$, i.e., it is complex. Results 
for $\rho_{ij}$ in the same systems of Figs.\ \ref{fermionization} 
and \ref{NOfermionization}(b) are presented in Fig.\ \ref{DensMatrix}.
Fig. \ref{DensMatrix}(a) shows that $|\rho_{ij}(\tau)|$ exhibits 
the same power-law decay than $\rho_{ij}$ in 
the ground state,\cite{rigol04_1} i.e., $|\rho_{ij}|\sim|x_i-x_j|^{-1/2}$
for large values of $|x_i-x_j|$ and {\it for all times}. Hence, 
this decay of the one-particle correlations 
is the one accounting for the large, and increasing, occupation of the 
lowest natural orbital as the system expands (like in the ground state).  
On the other hand, Figs.\ \ref{DensMatrix}(b)-(d) show that the phase 
of $\rho_{ij}$ ($\theta_{ij}$) starts to increasingly oscillate 
at large distances. This phase is the one accounting for the 
fermionization of $n_k$. In particular, Fig.\ \ref{DensMatrix}(b) 
shows that after a very short time, when the modulus of the OPDM have 
almost not changed, $\theta_{ij}$ has started to oscillate for 
$|x_i-x_j|\gg a$ producing a fast destruction of the zero momentum 
peak in $n^b_k$, as shown in Fig.\ \ref{fermionization}(a).

So far we have presented results for a trap with a characteristic 
density $\tilde{\rho}=0.51$ and $N_b=100$. In what follows we analyze 
the effects of changing $N_b$ and $\tilde{\rho}$ in the system. 
For that, we study the relative area between $n_k$ for HCB's and $n^f_k$ 
for fermions, $\delta=(\sum_k |n_k-n^f_k|)/(\sum_k n_k)$. We will 
consider that the $n_k$ of the HCB's has fermionized when 
$\delta\leq0.05$, i.e., when its difference with the one of the 
fermions is less or equal than a $5\%$. 

\begin{figure}[h]
\begin{center}
\includegraphics[width=0.93\textwidth,height=0.36\textwidth]
{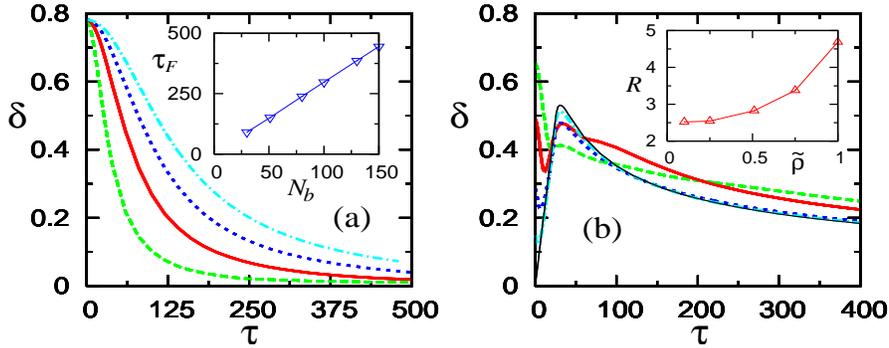} \vspace{-0.5cm}
\end{center}
\caption{Fermionization of $n_k$ during the expansion
of the gas. (a) Decrease of $\delta$ (see text) as a function of time 
for $\tilde{\rho}=0.51$; $N_b=51$ (dashed line), $N_b=100$ (continuous line), 
$N_b=150$ (dotted line), and $N_b=200$ (dashed-dotted line).
The inset shows the fermionization time $\tau_F$ (see text) vs $N_b$, 
for $\tilde{\rho}=0.51$.
(b) Decrease of $\delta$ as a function of time for $N_b=100$, 
$\tilde{\rho}=2.0$ (dashed line), $\tilde{\rho}=2.5$ (continuous line); 
and $N_b=101$, $\tilde{\rho}=3.0$ (dotted line), $\tilde{\rho}=4.0$ 
(dashed-dotted line). In the last two cases there is a Mott-insulating
domain in the center of the trap. We have plotted as a thin continuous
line $\delta$vs $\tau$ for $N_b=101$ in a pure Mott insulating state, 
i.e., a state with one particle per lattice site. The inset shows the
ratio $R$ between the size of gas when $\delta=0.05$ and its 
original size vs $\tilde{\rho}$ for $N_b=100$.}
\label{fermionizationvsrho}
\end{figure}

In Fig.\ \ref{fermionizationvsrho}(a) we show $\delta$ vs $\tau$ in 
systems with $\tilde{\rho}=0.51$ when the number of particles in the 
trap is increased. Notice that keeping $\tilde{\rho}$ constant is 
equivalent to keep constant the Fermi energy 
$\epsilon_F$.\cite{rigol03_3} Figure 
\ref{fermionizationvsrho}(a), and its inset, show that in this case the 
fermionization time ($\tau_F$) increases linearly with $N_b$. 
The fast disappearance of the $k=0$ peak in $n_k$ 
[Fig.\ \ref{fermionization}(a)] is reflected in 
Fig.\ \ref{fermionizationvsrho}(a) by a fast reduction of $\delta$. 
We find that after long times the reduction of $\delta$ is very close 
to a power law. This means that $\tau_F$ depends strongly on the 
criterion chosen.

The consequences of increasing $\tilde{\rho}$, and hence the Fermi 
energy, are analyzed in Fig.\ \ref{fermionizationvsrho}(b). 
In order to compare systems with different $\tilde{\rho}$, i.e., 
different $n_k$, we display in the inset of 
Fig.\ \ref{fermionizationvsrho}(b) the ratio $R$ 
between the size of the cloud when $\delta=0.05$, and its 
initial size (before expansion). (R is independent of the number 
of particles for a given value of $\tilde{\rho}$.\cite{rigol04_3}) 
The inset in Fig.\ \ref{fermionizationvsrho}(b) shows that with 
decreasing $\tilde{\rho}$ the ratio $R$ reduces up to $\sim 2.5$. 
For low $\tilde{\rho}$, such that the interparticle distance 
is much larger than the lattice spacing, the initial lattice 
gas is equivalent to the one in continuous systems. This means 
that $R\sim 2.5$ is relevant when there is no lattice, 
where the fermionization of $n_k$ can be more easily observed. 
In continuous systems, the dynamical fermionization of the HCB 
$n_k$ has been also recently studied.\cite{sutherland98,minguzzi05}
Interestingly, the fermionization time was found to be 
$\tau_F\sim 1/ \epsilon_F$,\cite{minguzzi05} in contrast to our 
findings in the lattice where it increases linearly with $N_b$ 
for a fixed $\epsilon_F$ [inset in Fig.\ \ref{fermionizationvsrho}(a)], 
and dramatically for large characteristic densities 
[Fig.\ \ref{fermionizationvsrho}(b)].

As shown in Fig.\ \ref{fermionizationvsrho}(b), increasing 
the characteristic density beyond the values depicted in its inset, 
the behavior of $\delta$ starts to depart from the one seen in 
Fig.\ \ref{fermionizationvsrho}(a). This is because particles 
become more localized in the middle of the trap, and after 
$\tilde{\rho}\sim 2.6-2.7$ a Mott insulator appears in the system.
This localization generates an $n_k$ that approaches the 
$n_k$ of the fermions in the initial state. [In the limit where all 
occupied lattice sites have $n_i=1$, $n_k(\tau=0)=n^f_k$.] 
When such systems are released from the trap $\delta$ increases
for some time. We will study in the next section the origin of this
increase. We should mention, however, that after long times 
a fermionization of $n_k$ starts to occur as before for smaller 
$\tilde{\rho}$. The difference is that, as shown in Fig.\ 
\ref{fermionizationvsrho}(b), the time scales for this process
are very large.

\section{Emergence of Quasi-Condensates at Finite Momentum\label{EQFM}}

We study in this section the free expansion of HCB's that are initially 
in a pure Mott-insulating state with one particle per lattice 
site. Since double occupancy is forbidden for very large on-site 
repulsions $U$, such states can be created experimentally in very deep 
optical lattices, and using strong confining potentials 
[$U\gg V_2(N_ba/2)^{2}\gg t$], so that vacancies in the initial state 
are suppressed.

\begin{figure}[h]
\begin{center}
\includegraphics[width=0.93\textwidth,height=0.35\textwidth]
{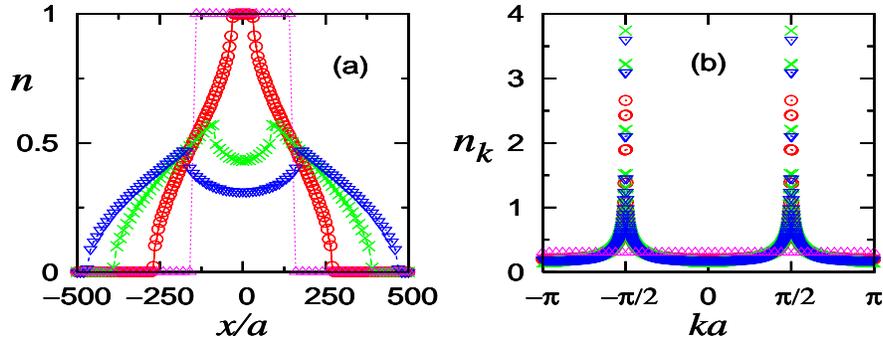} \vspace{-0.5cm}
\end{center}
\caption{Evolution of density (a) and momentum (b) 
profiles of 301 HCB's in 1000 lattice sites. The times are 
$\tau=0$ ($\triangle$), $60\hbar/t$ ($\bigcirc$), 
$120\hbar/t$ ($\times$), and $160\hbar/t$ ($\nabla$). Positions (a)
and momenta (b) are normalized by the lattice constant $a$.}
\label{perfil_fock}
\end{figure}

In Fig. \ref{perfil_fock} we show the evolution of density (a) and 
momentum (b) profiles of 301 HCB's after the confining potential 
is turned off. At $\tau=0$, $n_k$ is flat corresponding to a pure 
Fock state, where particles are localized in real space. The 
corresponding $n^f_k$ of the fermions is identical, and does not 
change with time. Surprisingly, sharp peaks appear in the HCB 
$n_k$ at $k=\pm \pi/2a$ during the expansion. Such peaks are very
similar to the ones present at $k=0$ in the ground state of a system
with superfluid domains.\cite{rigol04_1}

\begin{figure}[h]
\begin{center}
\includegraphics[width=0.93\textwidth,height=0.35\textwidth]
{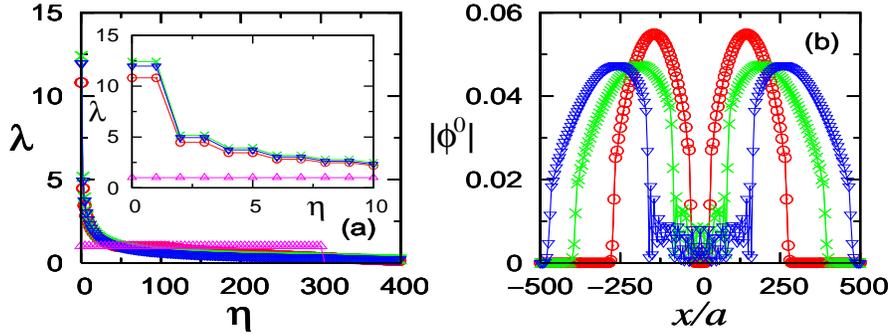} \vspace{-0.55cm}
\end{center}
\caption{Evolution of the natural orbital occupations (a) 
and the lowest natural orbital wave function (b) 
of 301 HCB's in 1000 lattice sites. The times are 
$\tau=0$ ($\triangle$), $60\hbar/t$ ($\bigcirc$), 
$120\hbar/t$ ($\times$), and $160\hbar/t$ ($\nabla$), 
and correspond to the density and momentum profiles shown 
in Fig.\ \ref{perfil_fock}.}
\label{NO_FOCK}
\end{figure}

Since the peaks at $n_{k=\pm \pi/2a}$ may indicate the presence of 
quasicondensates, like in the ground state, but at finite momentum
in this case, we diagonalize $\rho_{ij}$ and study the natural orbitals
and their occupations. In Fig.\ \ref{NO_FOCK}(a) we show the evolution
of the lowest 500 natural orbital occupations during the expansion.
Like in $n_k$, one can see that with the disappearance of the 
$\lambda=1$ plateau, which signals the Mott insulating state, a peak appears 
for the highest occupied orbitals. As seen in the inset of 
Fig.\ \ref{NO_FOCK}(a) (where we only show the eleven highest values 
of $\lambda$), the two highest occupied orbitals are degenerate. 
We discuss below that the two peaks seen in $n_k$ signal their 
emergence.

\begin{figure}[b]
\begin{center}
\includegraphics[width=0.93\textwidth,height=0.36\textwidth]
{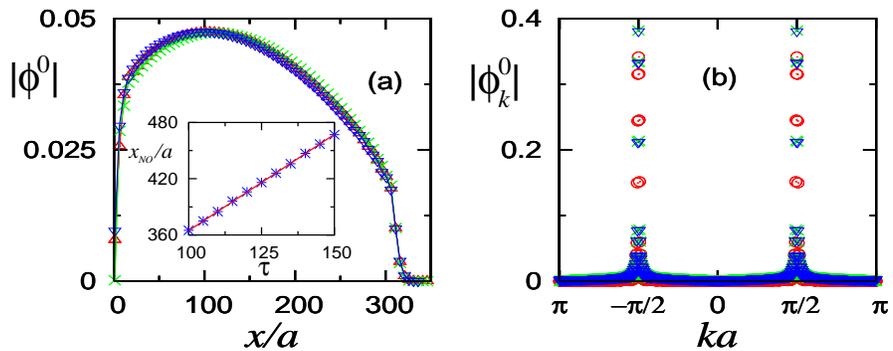} \vspace{-0.55cm}
\end{center}
\caption{(a) Superposed right lobe of the lowest natural 
orbital at $\tau=120\hbar/t$ ($\times$), 
$140\hbar/t$ ($\bigtriangleup$), and $160\hbar/t$ ($\nabla$). 
The inset shows the evolution of the lowest natural orbital 
right lobe's position. The line going through the points has a 
slope 2. (b) Fourier transform of the lowest natural orbital 
at $\tau=60\hbar/t$ ($\bigcirc$), $120\hbar/t$ ($\times$), 
and $160\hbar/t$ ($\nabla$).}
\label{NOK_FOCK}
\end{figure}
The evolution of the modulus of lowest natural orbital wave 
function ($\phi^0$ is complex) is shown 
in Fig.\ \ref{NO_FOCK}(b). The results for $|\phi^1|$ are
identical to those of $|\phi^0|$. (At $\tau=0$ the natural orbitals 
are delta functions at the occupied sites, so that we 
do not show $|\phi^0(\tau=0)|$ in the figure.) As seen in 
Fig.\ \ref{NO_FOCK}(b) during the expansion of the HCB's, 
two identical wave packets (or lobes) form at the sides of the 
Mott insulator. Once the Mott-insulating domain has disappeared, 
the shape of the wave packets almost does not change, like for a 
soliton, and they just move in opposite directions. 
To show that we have superposed in Fig.\ \ref{NOK_FOCK}(a) 
the right moving lobe at three different times. They are almost 
indistinguishable. 
This solitonic character of the moving wave packets allows us 
to calculate their group velocity, which we find to be constant
and equal to $v_{NO}=\pm 2at/\hbar$. This is shown in the inset 
of Fig.\ \ref{NOK_FOCK}(a) where we have plotted the position 
of the right moving wave packet $x_{NO}$ vs $\tau$, 
in units of $\hbar/t$, 
(the results for the left moving one are identical with 
$x\rightarrow -x$), which is a straight line with slope 2.   
Given the HCB dispersion relation in a lattice 
$\epsilon_k=-2t\cos ka$ (identical to the one of noninteracting 
fermions) one can obtain, using the well known expression 
$v=(1/\hbar)(\partial \epsilon_k/\partial k)$, the momenta 
corresponding to the moving lobes to be $k=\pm \pi/2a$. 

The above result can be further confirmed by calculating 
the Fourier transform of the lowest natural orbital at different 
times. They are depicted in Fig.\ \ref{NOK_FOCK}(b), and exhibit
sharp peaks centered at quasi-momenta $k=\pm \pi/2a$. 
The peak at $k=+\pi/2a$ appears due to the 
Fourier transform of the right lobe in Fig.\ \ref{NO_FOCK}(a) 
(i.e.\ $\phi^0_i \simeq |\phi^0_i| e^{i\pi x_i/2a}$ for $x_i>0$), 
and the one at $k=-\pi/2a$ due to the fourier transform 
of the left lobe (i.e.\ $\phi^0_i \simeq |\phi^0_i| e^{-i\pi x_i/2a}$ 
for $x_i<0$). We have also calculated the Fourier transform
of the other natural orbitals finding that they have 
a very small or zero weight at $k=\pm \pi/2a$. Hence, 
the peaks appearing at $n_{k=\pm \pi/2a}$ are reflecting 
the existence of the two traveling, with $k=\pm \pi/2a$, 
highly occupied natural orbitals.

\begin{figure}[h]
\begin{center}
\includegraphics[width=0.93\textwidth,height=0.36\textwidth]
{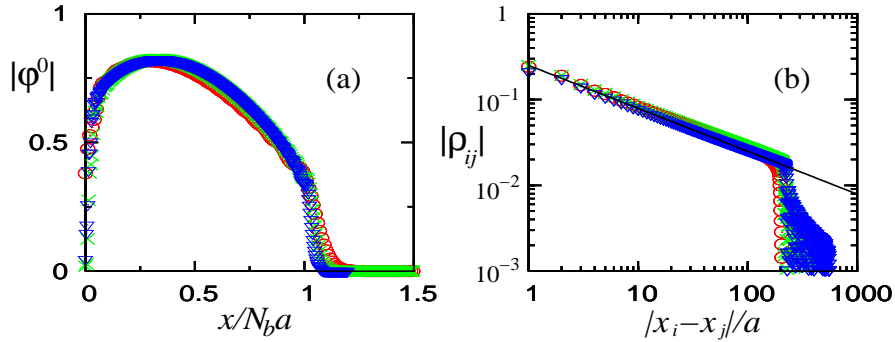} \vspace{-0.55cm}
\end{center}
\caption{(a) Scaled right lobe of the lowest natural orbital (see text) 
vs $x/N_ba$ for $N_b=101$ ($\bigcirc$), $201$ ($\times$), 
and $301$ ($\nabla$). (b) One-particle density matrix measured with 
$x_i$ fixed at the left extreme of the left lobe, and $x_j>x_i$. 
Times are the same of Figs.\ \ref{perfil_fock}-\ref{NOK_FOCK},
$\tau$=$60\hbar/t$ ($\bigcirc$), $120\hbar/t$ ($\times$), and 
$160\hbar/t$ ($\nabla$). The straight line is $0.25\ |(x_i-x_j)/a|^{-1/2}$ 
(see text).}
\label{DensMatFock}
\end{figure}

Apart from their large occupations, the two lowest natural orbitals
exhibit another remarkable property. Once formed, their moving 
lobes have exactly the same form independent of the number of 
particles ($N_b$) in the initial Mott insulating state. Their 
size $L$ is proportional to $N_b$. Actually, as shown in 
Fig.\ \ref{DensMatFock}(a) the lobes can be perfectly rescaled 
considering $|\varphi_0|=N_b^{1/2}|\phi^0|$ vs $x/N_b a$.

In order to understand the origin of the large occupation of the 
two lowest natural orbitals, and their scaling properties, we 
study the one-particle density matrix. At $\tau=0$, in the Mott 
insulating state, $\rho_{ij}=\delta_{ij}$ at the occupied lattice 
sites, and zero in the rest of the system. During the expansion 
of the HCB's, a surprising behavior sets in. Quasi--long-range
one-particle correlations develop in the region of the system 
where the lobes of the lowest two natural orbitals were observed
in Figs.\ \ref{NO_FOCK}(b)-\ref{DensMatFock}(a). The decay of 
these correlations is $|\rho_{ij}|=0.25\ |(x_i-x_j)/a|^{-1/2}$, 
as depicted in Fig.\ \ref{DensMatFock}(b). They exhibit the same 
power-law decay that was shown to be universal, independent 
of the exponent of the trapping potential, in the ground 
state.\cite{rigol04_1} Outside the regions where the lobes are
present, $\rho_{ij}$ decays faster as seen in the tails of 
$|\rho_{ij}|$ present in Fig.\ \ref{DensMatFock}(b). 
These quasi--long-range correlations in 
two disconnected segments of the system are the reason for the 
degeneracy found in the lowest natural orbitals. A detailed 
analysis of $\rho_{ij}$ also shows that, as expected, the peak 
in $n_k$ at $k=+\pi/2a$ is originated by components of 
$\rho_{ij}$ with $x_i,x_j>0$, and the one at $k=-\pi/2a$ by 
the components with $x_i,x_j<0$, so that in the regions of the 
lobes one can write
$\rho_{ij}\simeq 0.25\ |(x_i-x_j)/a|^{-1/2}e^{i\pi(x_i-x_j)/2a}$ 
for $x_i,x_j>0$, and 
$\rho_{ij}\simeq 0.25\ |(x_i-x_j)/a|^{-1/2}e^{-i\pi(x_i-x_j)/2a}$ 
for $x_i,x_j<0$. The prefactor of the power law (0.25) was 
found to be independent of time, and of the number of particles 
in the initial Fock state.

\begin{figure}[h]
\begin{center}
\includegraphics[width=0.93\textwidth,height=0.36\textwidth]
{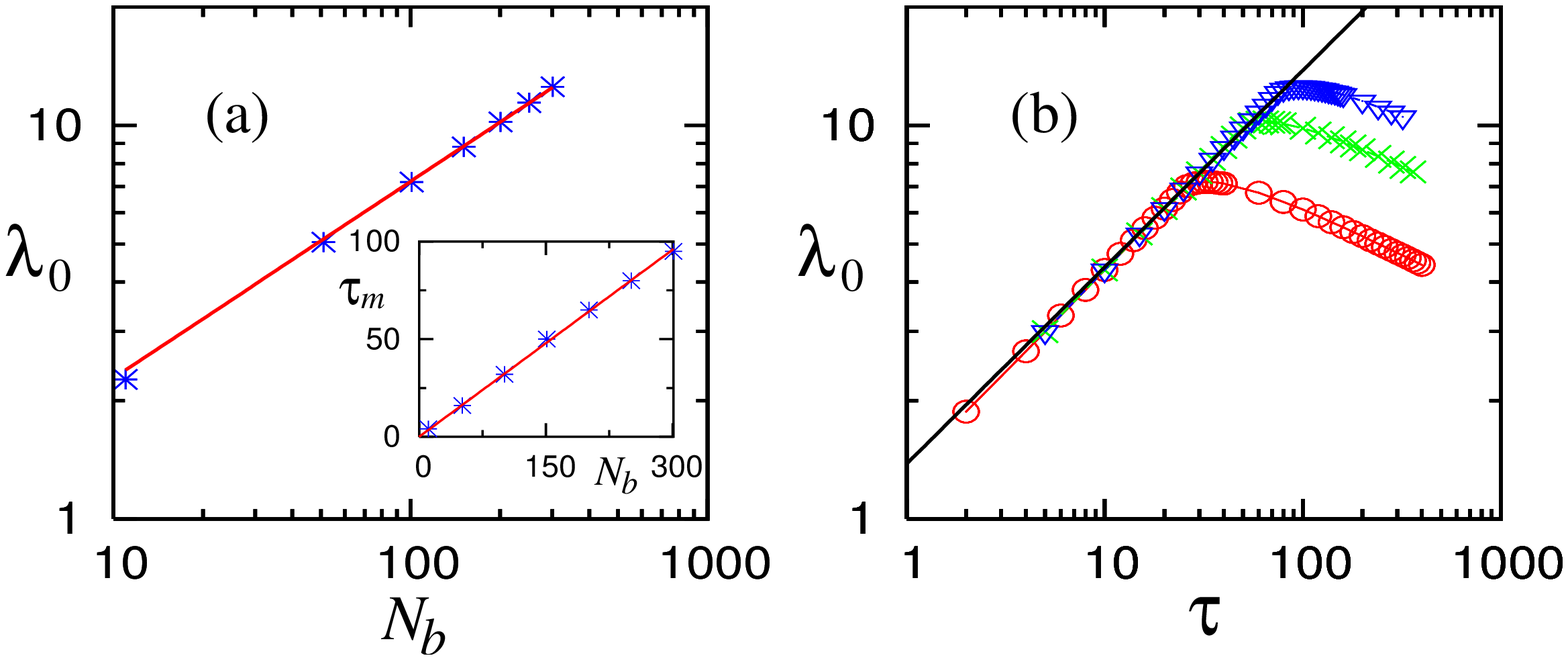} \vspace{-0.55cm}
\end{center}
\caption{(a) Maximum occupation of the lowest natural 
orbital vs $N_b$. The straight line is $0.72\ N_b^{1/2}$.
The inset shows the time at which the maximum occupation of 
$\lambda_0$ is reached $\tau_m$ (in units of $\hbar/t$) 
vs $N_b$. The line following the points is $0.32N_b$.
(b) Time evolution of the lowest NO occupation for 
$N_b=101$ ($\bigcirc$), $201$ ($\times$), and $301$ ($\nabla$). 
The straight line is $1.38\sqrt{\tau t/\hbar}$.}
\label{PowerLaws}
\end{figure}

Considering all the properties discussed so far one can show that
the two highest occupied natural orbitals, dynamically generated 
and traveling with momenta $\pm \pi/2$, are quasicondensates 
in the same sense than those observed in the ground state.\cite{rigol04_1}
They have a diverging occupation $\lambda_0$ as the number of particles 
in the initial Mott insulating state increases, however 
$\lambda_0/N_b\rightarrow 0$ as $N_b\rightarrow \infty$. 
$\lambda_0$ can be obtained as
\begin{eqnarray}
\lambda_0&=&\sum_{ij}\phi^{*0}_i \rho_{ij}\phi^0_j\simeq 
2/a^2 \int^L_{0}dx \int^L_{0}dy \ \phi^{*0}(x)\rho(x,y) \phi^0(y)
\nonumber\\ &=& N_b^{1/2} \int^{1}_{0}dX \int^{1}_{0}dY 
\frac{|\varphi^0(X)|0.25|\varphi^0(Y)|}{|X-Y|^{1/2}} 
=A\sqrt{N_b}. 
\label{lambda0}
\end{eqnarray}
The sums were replaced by integrals considering that $L\gg a$. 
In the second line of Eq.\ (\ref{lambda0}) we changed  
variables $x$=$X N_ba$, $y$=$Y N_ba$, $\phi^0$=$N_b^{-1/2}\varphi^0$, 
and noticed that the phase factors between the natural orbital and 
$\rho_{ij}$ cancel out. The integral over $X,Y$ is a constant that 
we call $A$. The final result in Eq.\ (\ref{lambda0}) shows that, 
like in the ground state, $\lambda_0\sim \sqrt{N_b}$. 
This result was confirmed by our numerical calculations as
shown in Fig.\ \ref{PowerLaws}(a). There we have plotted the 
maximum occupation of the lowest natural orbital vs $N_b$. 
A fitting of the constant $A$ reveals a value of 0.72.

The appearance of two traveling quasicondensates, 
at momenta $k=\pm \pi/2a$, can be understood on the basis 
of total energy conservation. Given the HCB dispersion relation 
in a lattice $\epsilon_k=-2t\cos ka$, we notice that the 
total energy of the initial Mott insulating state is zero 
($E_T = 0$), since $n_k$ is totally flat. Considering that 
the energy is conserved in the system, if all the particles 
would condense into one state, it would be the one with an 
energy corresponding to $\overline{\epsilon}_k = E_T/N=0$.
For the case of free expanding HCB's there are two states
with $\overline{\epsilon}_k=0$. They are the momentum 
states with $k=\pm \pi/2a$. Actually, in 1D there is only
quasicondensation, so that the argument above applies only 
to maximize the occupation of the $k=\pm \pi/2a$ states. 
Two other important properties of the $k=\pm \pi/2a$ wave 
packets are that they travel with the maximum velocity possible 
in a lattice, and that the density of states of the free 
system has minima at these values of $k$. The latter property 
reinforces quasicondensation into the two single 
$k=\pm \pi/2a$ states.

During the formation of the quasicondensates, the increase
of their occupation is also characterized by a power law. 
This is shown in Fig.\ \ref{PowerLaws}(b), where we plot
$\lambda_0$ as a function of the evolution time. 
The {\it log-log} scale evidences the fast increase of 
$\lambda_0$. A detailed examination shows that the 
population of the quasicondensate increases in a 
universal way [$1.38\sqrt{\tau t/\hbar}$, continuous line in 
Fig.\ \ref{PowerLaws}(b)] independently of the initial number 
of particles in the Fock state. This power law is determined 
by the universal behavior of the off-diagonal part of $\rho_{ij}$ 
shown before, and by the fact that during the formation of the 
quasicondensates, they increase their sizes linearly with time 
at a rate given by the maximum velocity in the lattice 
$|v_{NO}| = 2at/\hbar$. The power law mentioned before 
is followed up to the point where the maximal occupation is 
reached, which corresponds to the point at which the initial 
Mott insulator has completely melted. One can then realize that 
given $v_{NO}$, the time at which this occurs $\tau_m$ depends 
linearly on the number of particles. We find that 
$\tau_m=0.32N_b\hbar /t$, as shown in the inset of 
Fig.\ \ref{PowerLaws}(a). To have an example of the time scales 
we are referring here, we consider a tipical experimental setup 
of rubidium atoms, in a lattice with a recoil energy of $E_r=20$kHz, 
and a depth of 20$E_r$. In these systems, the time $\tau_m$ can 
be estimated as $\tau_m\sim 5.7N_b$(ms).

A second characteristic of these systems that can be seen 
in the {\it log-log} plot in Fig.\ \ref{PowerLaws}(b) is the 
slow reduction of the lowest natural orbital occupation, 
with the consequent reduction of
the $n_{k=\pm \pi/2}$ peaks. This slow reduction is consistent 
with the slow fermionization of $n_k$ seen in 
Fig.\ \ref{fermionizationvsrho}(b), and suggest that such an 
arrangement could be used to create atom lasers with a wavelength 
that can be fully controlled given the lattice parameter $a$. 
No additional effort is needed to separate the quasi-coherent 
part from the rest since the quasicondensate is traveling at 
the maximum velocity on the lattice so that the front 
part of the expanding cloud is the quasi-coherent part.

\begin{figure}[h]
\begin{center}
\includegraphics[width=0.93\textwidth,height=0.34\textwidth]
{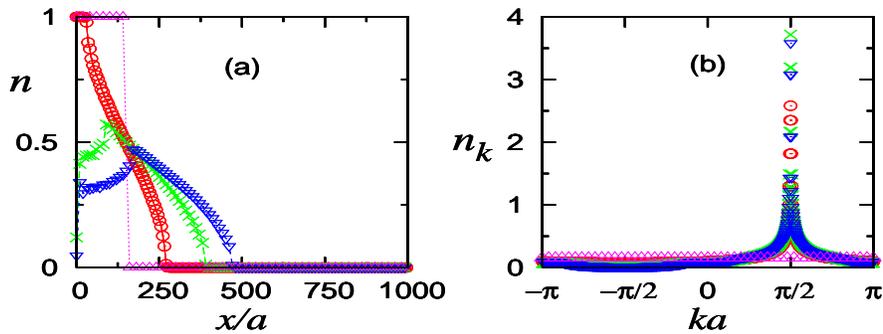} \vspace{-0.5cm}
\end{center}
\caption{Evolution of density (a) and momentum (b) 
profiles of 150 HCB's evolving only to the right 
(the system is open in the left extreme) in 1000 lattice sites. 
The times are $\tau=0$ ($\triangle$), $60\hbar/t$ ($\bigcirc$), 
$120\hbar/t$ ($\times$), and $160\hbar/t$ ($\nabla$).}
\label{LASER}
\end{figure}

The actual experimental realization would imply to restrict 
the evolution of the initial Mott state to one direction, 
as shown in Fig.\ \ref{LASER}. There we display the density and
momentum profiles of 150 HCB's restricted to evolve to the right 
in 1000 lattice sites at the same evolution times of Fig.\ 
\ref{perfil_fock}. This figure shows that the values of 
$n_{k=\pi/2}(\tau)$ are almost the same than the ones in 
Fig.\ \ref{perfil_fock}, although the initial Fock state 
has half the number of particles. This is because in this 
case only one quasicondensate is created, so that 
the lowest natural orbital also has the same occupation
than in Fig.\ \ref{NO_FOCK}(a), and is not degenerate 
anymore.

The above results suggest how to proceed 
to obtain lasers in higher dimensional systems 
where real condensation can occur.\cite{lieb02}
One could employ Mott-insulating states with one particle 
per lattice site created by a very strong on-site repulsive 
potential $U$. The latter is required in order to obtain a 
close realization of a pure Fock state, since quantum 
fluctuations of the particle number present in a Mott 
insulator for any finite $U$ will be strongly 
suppressed.\cite{batrouni02,rigol03_1,rigol03_2} 
Then the geometry of the lattice should be designed in order 
to restrict the evolution of the Mott insulator to one 
direction only, and to have a low density of states around 
the mean value of energy per particle in the initial state. 
With these conditions the sharp features observed in 1D 
may be reproduced by a condensate in higher dimensions.

\section{Information Entropies\label{IE}}

We study in this section the Shannon information 
entropies\cite{shannon48} in coordinate ($S_n$) and momentum ($S_k$) 
space for the two physical situations discussed 
in Secs.\ \ref{DF} and \ref{EQFM}, and in between.

Here we will follow the definitions
\begin{equation}
S_n=-\sum_i \overline{n}_i \ln \overline{n}_i,
\end{equation} 
and
\begin{equation}
S_k=-\sum_k \overline{n}_k \ln \overline{n}_k,
\end{equation}
normalizing $\overline{n}_i$ and $\overline{n}_k$ such that 
$\sum_i \overline{n}_i=\sum_i \overline{n}_i=1$. Since we have 
studied the expansion of HCB's in systems with $\sim 2000$ lattice 
sites, in all the cases discussed in this section we discretize 
the Brillouin zone with 2000 points to calculate $S_k$.

In continuous systems, the sum of $S_n$ and $S_k$ satisfies several
interesting properties. It was rigorously proven by Bialynicki-Birula 
and Mycielski\cite{birula79} that when density and momentum 
distributions are normalized to unity,
\begin{equation}
S_T=S_n+S_k\geq d(1+\ln \pi),
\end{equation}
with $d$ the space dimension. At the Hartree-Fock level, it has been
shown that in the ground state of atoms $S_T$ acquires its minimum 
value.\cite{grade85} And more recently, it has been proposed that 
$S_T$ can be used as a measure of correlation in atomic 
systems.\cite{guevara03}

In what follows, the comparison of $S_n$ and $S_k$ with their 
fermionic counterparts defined in terms of the fermionic densities 
($S^f_n=-\sum_i \overline{n}^f_i \ln \overline{n}^f_i$) and momentum 
distributions ($S^f_k=-\sum_k \overline{n}^f_k \ln \overline{n}^f_k$), 
will be shown to be very useful. 
Since density profiles are identical for HCB's and noninteracting 
fermions, at any time, $S_n(\tau)$ and $S^f_n(\tau)$ are the same. 
This identity does not hold in momentum space. In the ground state, 
given the large occupation of the low momentum states in the TG gas, 
in contrast to the extended $n_k$ of the fermions, one can realize that 
$S^f_k(\tau=0)\geq S_k(\tau=0)$, where the equality only holds in 
the limit in which HCB's and fermions are localized in real space 
(the systems in Sec.\ \ref{EQFM}). 

\begin{figure}[h]
\begin{center}
\includegraphics[width=0.93\textwidth,height=0.36\textwidth]
{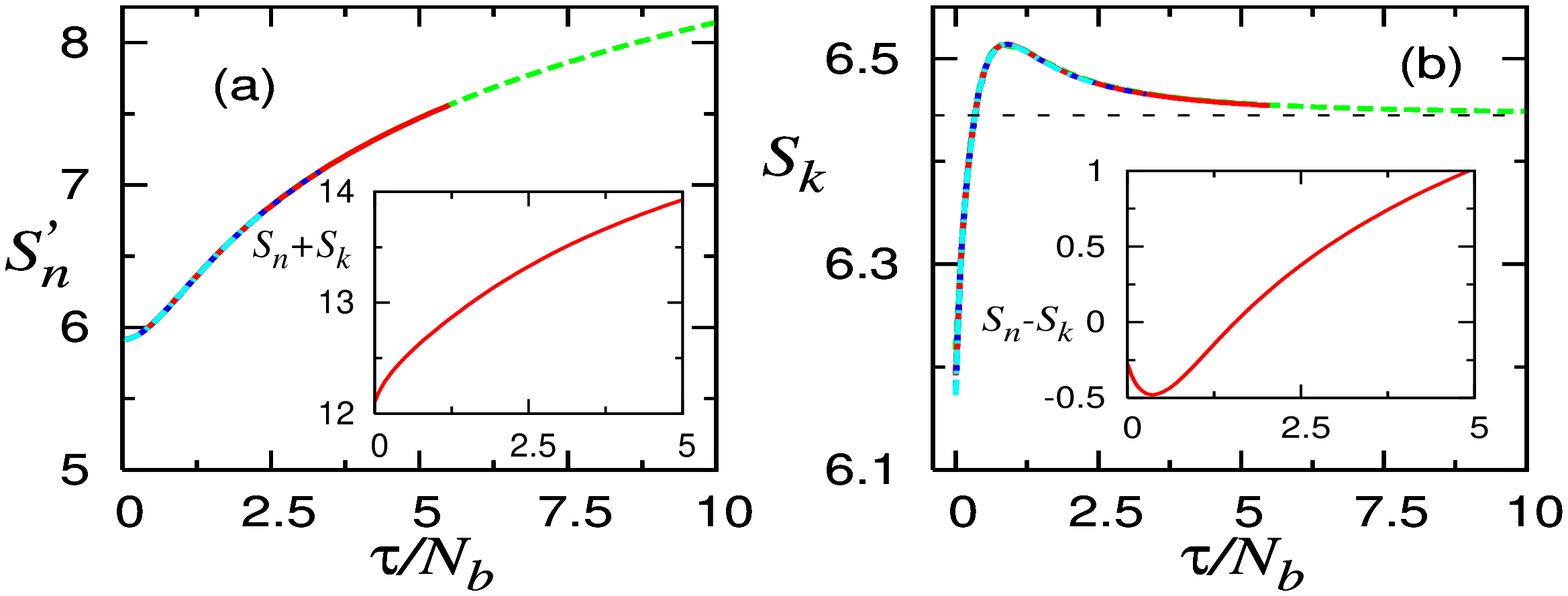} \vspace{-0.5cm}
\end{center}
\caption{Evolution of $S'_n$ (a) and $S_k$ (b) in systems with
2000 lattice sites. The initial state has $\tilde{\rho}=1.0$
in all cases, and $N_b=51$ (dashed line), $N_b=100$ (continuous line), 
$N_b=150$ (dotted line), and $N_b=200$ (dashed-dotted line). 
In (a) we have displaced the $S_n$ of $N_b=51$, $N_b=150$, 
and $N_b=200$ by a constant (see text), to superpose it with the one of 
$N_b=100$. The insets in (a) and (b) show the sum and difference,
respectively, of $S_n$ and $S_k$ for $N_b=100$. The thin dashed line 
in (b) shows $S^f_k$, which does not change in time.}
\label{EntropyvsNb}
\end{figure}

We first study the behavior of $S_n$ and $S_k$ for the cases in which 
a fast fermionization of $n_k$ was observed in Sec.\ \ref{DF}. In 
Fig.\ \ref{EntropyvsNb}(a) we show the evolution of $S_n$ for systems
with the same characteristic density and different $N_b$, like in 
Fig.\ \ref{fermionizationvsrho}(a). As the number of particles in the 
trap is increased keeping $\tilde{\rho}$ constant, 
the initial system size increments proportionally 
to $N_b$, producing an increase of $S_n$. In order to compare $S_n$ 
for different number of HCB's we have displaced them to coincide
with $S_n$ for $N_b=100$. This can be easily achieved considering
$S'_n=S_n+\ln(100/N_b)$. Fig.\ \ref{EntropyvsNb}(a) shows that 
rescaling $\tau$ by the number of particles, the time evolution
of the different systems collapse over the same curve which, as 
expected, shows an increase of $S_n$ as the system expands.

The behavior of $S_k$, for the systems in Fig.\ \ref{EntropyvsNb}(a),
is shown in Fig.\ \ref{EntropyvsNb}(b). Since the characteristic density 
is kept constant, at $\tau=0$, $S_k$ for different $N_b$ has the same
value. During the expansion one can see that $S_k$ overcomes $S^f_k$ 
(that does not change since the fermions are noninteracting).
This means that, taking into account that the Shannon entropies in 
coordinate space are identical for HCB's and fermions, as the TG gas 
expands $S_T$ of the HCB's becomes larger than the corresponding one 
$S^f_T$ of the fermions. For large times $S_T\rightarrow S^f_T$ from 
above. As discussed below, the increase of $S_T$ beyond the value of 
$S^f_T$ is a characteristic of the fermionization process studied in 
Sec.\ \ref{DF}.

We have also plotted in the insets of Figs.\ \ref{EntropyvsNb}(a) and 
(b) the sum and differences, respectively, of $S_n$ and $S_k$. The sum
shows that $S_T$ has its minimum value in the ground state. Hence, the 
expansion only produces its increment. The difference $S_n-S_k$ in 
the inset of Figs.\ \ref{EntropyvsNb}(b) reflects the fast destruction 
of the $n_{k=0}$ peak at very short times, which produces changes in 
$S_k$ that are larger than the ones in $S_n$, so that for short times
$S_k$ has the largest contribution to the increase of $S_T$. For large 
expansion times $S_n$ is the dominant contribution to $S_T$ as 
$S_k\rightarrow S^f_k$ slowly.

The properties of the information entropies that we have discussed so far
are generic for all systems that start their evolution from a state
with low density in the lattice, and in which a fast fermionization of 
$n_k$ occurs. Hence, we infer that they will also be valid in the 
continuum case,\cite{kinoshita04,sutherland98,minguzzi05} which can be
obtained as the low density limit in the lattice.\cite{rigol04_1} 
As the characteristic density in the system increases, and a Mott 
insulator appears in the middle of the trap, the behavior of $S_k$ and 
the way in which $S_k\rightarrow S^f_k$ change qualitatively.

\begin{figure}[h]
\begin{center}
\includegraphics[width=0.93\textwidth,height=0.36\textwidth]
{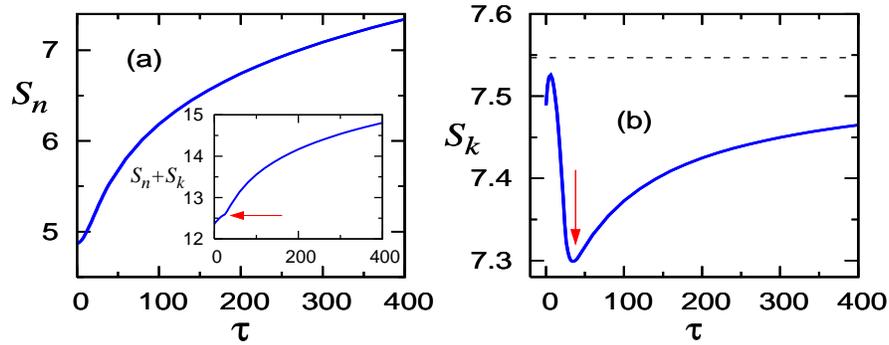} \vspace{-0.5cm}
\end{center}
\caption{Evolution of $S_n$ (a) and $S_k$ (b) in a system with
101 HCB's, 2000 lattice sites, and initial $\tilde{\rho}=3.0$. 
In the inset we show the sum of $S_n$ and $S_k$. 
The thin dashed line in (b) shows $S^f_k$.}
\label{EntropyMott}
\end{figure}

\begin{figure}[b]
\begin{center}
\includegraphics[width=0.93\textwidth,height=0.35\textwidth]
{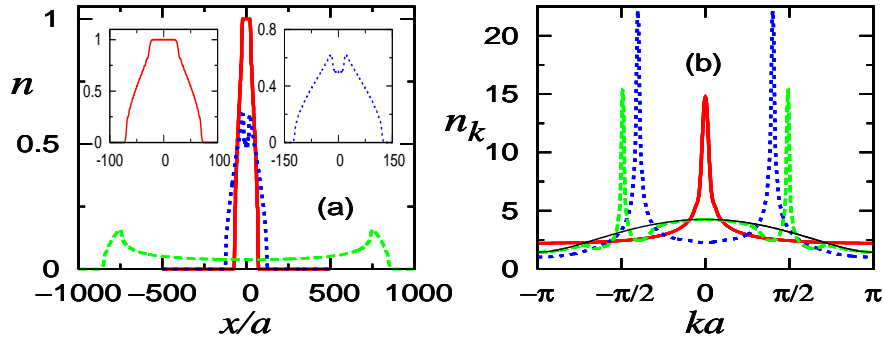} \vspace{-0.5cm}
\end{center}
\caption{Evolution of density (a) and momentum (b) 
profiles of 101 HCB's in 2000 lattice sites, for and 
initial $\tilde{\rho}=3.0$. The times are 
$\tau=0$ (continuous line), $34\hbar/t$ (dotted line), 
$400\hbar/t$ (dashed line). The density profiles at 
$\tau=0$ and $\tau=34\hbar/t$ can be better seen in the 
insets in (a). In (b) we have plotted as a thin continuous line
the momentum distribution function of the fermions.}
\label{Perfil_Mott}
\end{figure}

In Fig.\ \ref{EntropyMott}, we show the time evolution of $S_n$
and $S_k$ for a system with $\tilde{\rho}=3.0$, which at $\tau=0$
has a Mott insulating domain in the middle of the trap 
[Fig.\ \ref{Perfil_Mott}(a)]. During the expansion of the 
HCB's for this initial condition $S_n$ [Fig.\ \ref{EntropyMott}(a)]
is very similar to the one in Fig.\ \ref{EntropyvsNb}(a) for low
initial densities, i.e., no distinguishing feature appears
in $S_n$ for different initial densities. 

On the contrary, the behavior of $S_k$ in Fig.\ \ref{Perfil_Mott}(b)  
is very different to the one seen in Fig.\ \ref{EntropyvsNb}(b),
with the exception of very short times where $S_k$ in 
Fig.\ \ref{Perfil_Mott}(b) exhibits a small increase. It corresponds 
to the fast reduction of the $n_{k=0}$ peak similarly to the one seen
in Fig.\ \ref{fermionization}(a). At $\tau=0$, there is a peak at 
$k=0$ [Fig.\ \ref{Perfil_Mott}(b)] because of the superfluid domains
surrounding the Mott insulator in the middle of the trap [left inset 
in Fig.\ \ref{Perfil_Mott}(a)]. At short times these superfluid 
regions are affected by the expansion like the systems analyzed in 
Sec.\ \ref{DF}, producing an increase of $S_k$.  

With the melting of the Mott insulator in the middle of 
the trap, quasi-coherence builds between initially uncorrelated 
particles producing two peaks in $n_k$. As shown in 
Fig.\ \ref{Perfil_Mott}(b) for $\tilde{\rho}=3$, they are even 
larger than the one in the ground state. This causes the 
reduction of $S_k$ seen in Fig.\ \ref{EntropyMott}(b). Notice that
the situation is similar to the one discussed in Sec.\ \ref{EQFM}, 
but now the initial condition is different since the Mott insulator 
is surrounded by superfluid phases. Because of that, the peaks in 
$n_k$ do not appear at $k=\pm \pi/2$ [Fig.\ \ref{Perfil_Mott}(b)], 
although with time they move toward $k=\pm \pi/2$. In this case 
the lowest natural orbitals are, during the expansion, more complicated 
than the ones described in Sec.\ \ref{EQFM} since HCB's with many 
different momenta are couple into them. Hence, in experiments, the 
quality of the $k=\pm \pi/2$ traveling quasicondensates needs to be 
ensured by reducing the size of the superfluid domains surrounding 
the Mott-insulating core.

After the peaks in $n_k$ have acquired their maximum value, $S_k$
starts to slowly increase towards $n_k$. As shown in the inset of
Fig.\ \ref{EntropyMott}(a) the sum of $S_n$ and $S_k$ always 
increases during the expansion. Even when $S_k$ decreases, 
the increase of $S_n$ is larger and warrants that the minimum 
of $S_T$ occurs in the ground state. As signaled by arrows 
in Fig.\ \ref{EntropyMott}, a kink can be observed in $S_T$ when 
$S_k$ attains its minimum value.

\begin{figure}[h]
\begin{center}
\includegraphics[width=0.93\textwidth,height=0.36\textwidth]
{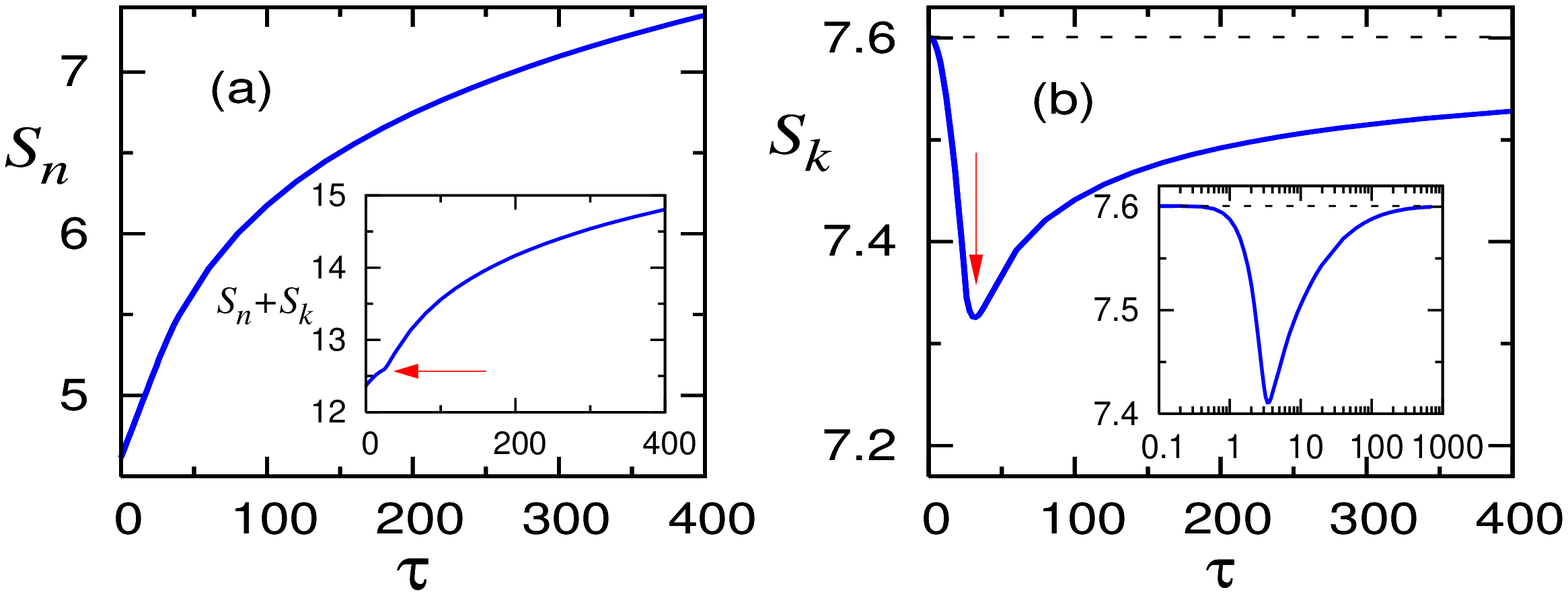} \vspace{-0.5cm}
\end{center}
\caption{Evolution of $S_n$ (a) and $S_k$ (b) in a system with
101 HCB's, 2000 lattice sites, and initially in a pure Mott 
insulating state. In the inset in (a) we show the sum of 
$S_n$ and $S_k$. In the inset in (b) we have plotted $S_k$ vs 
$\tau$ for an initial Mott insulating state with 11 HCB's and 
expanding in a system with up to 3000 lattice sites. 
The thin dashed lines show $S^f_k$.}
\label{EntropyFock}
\end{figure}

We conclude this section with the analysis of the information 
entropies when the system starts its evolution from pure 
Mott-insulating states, like the ones discussed in Sec.\ \ref{EQFM}.
As shown in Figs.\ \ref{EntropyFock}(a),(b), and the inset in 
(a), the behavior of $S_n$ and $S_k$ is the same 
observed in Fig.\ \ref{EntropyMott} [if one excludes the very 
short times when $S_k$ increases in Fig.\ \ref{EntropyMott}(b)
due to the superfluid domains that are not present in the case 
presented in Fig.\ \ref{EntropyFock}]. Since $S_k$ in these
systems is always smaller than $S^f_k$, we have found a criterion 
in terms of the Shannon information entropy in momentum space
that allows to distinguish between (i) systems in which a fast 
fermionization of $n_k$ occurs (Sec.\ \ref{DF}) where 
$S_k\rightarrow S^f_k$ from above, and (ii) systems in which
quasi-coherence develops dynamically producing traveling 
quasicondensates (Sec.\ \ref{EQFM}) where 
$S_k\rightarrow S^f_k$ from below. The latter feature is more 
difficult to see for the number of particles we have employed 
all over this work ($N_b=100$--300), due to the large system 
sizes one requires to calculate. However, in the inset in 
Fig.\ \ref{EntropyFock}(b) we present results obtained
for the expansion of 11 HCB's in up to 3000 lattice sites 
were one can clearly see that $S_k$ is always smaller than 
$S^f_k$ and that $S_k\rightarrow S^f_k$ for very long times.

\section{Conclusions\label{C}}

\vspace{0.3cm}
In this work we have presented a unified study of the expansion
of TG gases in 1D optical lattices.\cite{rigol04_3,rigol04_2} 
We have employed an exact numerical
approach that allows to study systems with large number of particles 
and sizes, and for arbitrary long time scales. We find that two very 
different regimes can be observed according to the initial conditions 
of the trapped gas.

(Regime I) When the TG gas starts it free expansion from a low 
density system, the momentum distribution of the bosons
rapidly approaches the one of the equivalent noninteracting fermions.
This fermionization of $n_k$ occurs without loss of coherence 
in the system. Actually, coherence increases as shown by the 
increase of the occupation of the lowest natural orbitals. This can 
be understood due to the presence of quasi-long range one-particle 
correlations, which have the same power-law decay as in the 
equilibrium case. A new feature that appears during the expansion 
of the gas is that the lowest natural orbital starts to be populated 
by particles with many different momenta, in contrast to the ground 
state where it is mainly populated by particles with $k=0$.   

(Regime II) When the TG gas starts it free expansion from a high 
density (Mott-insulating) trapped system, quasi-long range 
correlations develop between initially uncorrelated particles.
They exhibit the same power-law decay known from the ground state,
and produce the emergence of quasicondensates of HCB's at finite 
momentum. Their maximum occupation was found to be proportional to 
the squared root of the number of particles in the system. 
This quasi-condensation out-of-equilibrium is reflected by the 
appearance of two identical peaks in the momentum distribution 
function. We find that the dynamically generated quasicondensates 
are very similar to solitons since they can travel long distances 
almost without changing the shape of the wave-function and 
their occupation. Their momenta can be fully controlled by means 
of the lattice parameter.

Finally, we have shown that the Shannon information entropy in 
momentum space provides a criterion to distinguish the two regimes
previously mentioned. In the first regime the Shannon information 
entropy in momentum space becomes larger than the corresponding 
one of the fermions, and with time approaches the last one from 
above. For the second regime we find that the Shannon information 
entropy in momentum space is always smaller for HCB's than for 
fermions so that the first one approaches the second one from below.
(The fermionic entropy in momentum space remains unchanged during 
the free expansion since they are noninteracting particles.)
The information entropy in momentum space was also shown to be
useful to study the crossover between both regimes. It allows 
to distinguish the times scales in which the different processes
related to the superfluid and Mott-insulating domains set in 
the momentum distribution function.

\section{Acknowledgments}

We are grateful to R. T. Scalettar and R. R. P. Singh for 
stimulating discussions. This work was supported 
by NSF-DMR-0312261, NSF-DMR-0240918, NSF-ITR-0313390, 
and SFB 382. We thank HLR-Stuttgart (Project DynMet) 
for allocation of computer time.

\end{document}